\documentclass[a4paper,11pt]{article}

\usepackage[hang,flushmargin]{footmisc} 

\usepackage{authblk}

\usepackage[english]{babel}
\usepackage[utf8]{inputenc}
\usepackage{amsmath}
\usepackage{graphicx}
\usepackage[colorinlistoftodos]{todonotes}
\usepackage[square]{natbib}
\usepackage{tikz}
\usepackage{hyperref}
\hypersetup{
	colorlinks,
	citecolor=black,
	filecolor=black,
	linkcolor=black,
	urlcolor=black
}
\usepackage{geometry}
\makeatletter
\renewcommand\section{\addtocontents{toc}{\protect\addvspace{0\p@}}
  \@startsection {section}{1}{\z@}%
  {-3.5ex \@plus -1ex \@minus -.2ex}%
  {2.3ex \@plus.2ex}%
  {\normalfont\Large\bfseries}}
\makeatother

\usepackage{ifthen}

\usepackage{fancyhdr}
\fancypagestyle{firststyle}
{
  \fancyhf{}%
  \fancyhead[L]{\footnotesize Chapter pre-print to appear in the \textit{Oxford Handbook on Algorithmic Music. Oxford University Press}}
}
\title{\textbf{The Machine Learning Algorithm as Creative Musical Tool}}

\author{Rebecca Fiebrink}
\author{Baptiste Caramiaux}
\affil{\small Goldsmiths College, University of London\\
New Cross, London, UK\\
\{r.fiebrink,b.caramiaux\}@gold.ac.uk}

\date{}

\begin{document}

\maketitle

\thispagestyle{firststyle}
\pagestyle{fancy}

\begin{abstract}

Machine learning is the capacity of a computational system to learn structures from datasets in order to make predictions on newly seen data. Such an approach offers a significant advantage in music scenarios in which musicians can teach the system to learn an idiosyncratic style, or can break the rules to explore the system's capacity in unexpected ways. In this chapter we draw on music, machine learning, and human-computer interaction to elucidate an understanding of machine learning algorithms as creative tools for music and the sonic arts. We motivate a new understanding of learning algorithms as human-computer interfaces. We show that, like other interfaces, learning algorithms can be characterised by the ways their affordances intersect with goals of human users. We also argue that the nature of interaction between users and algorithms impacts the usability and usefulness of those algorithms in profound ways. This human-centred view of machine learning motivates our concluding discussion of what it means to employ machine learning as a creative tool.

\end{abstract}

\section{Introduction}

Machine learning algorithms lie behind some of the most widely used and powerful technologies of the 21st century so far. Accurate voice recognition, robotics control, and shopping recommendations stand alongside YouTube cat recognisers \citep{le2013building} as some of machine learning's most impressive recent achievements. Like other general-purpose computational tools, machine learning has captured the imaginations of musicians and artists since its inception. Sometimes musicians politely borrow existing machine learning algorithms and use them precisely as they were intended, providing numerous well-chosen examples of some phenomenon and then using an appropriate algorithm to accurately model or recognize this phenomenon. Other times, musicians break the rules and use existing algorithms in unexpected ways, perhaps using machine learning not to accurately model some phenomenon implicit in the data but to discover new sounds or new relationships between human performers and computer-generated processes. In still other cases, music researchers have formulated their own new definitions of what it means for a computer to learn, and new algorithms to carry out that learning, with the specific aim of creating new types of music or new musical interactions.

\subsection{What is this Chapter?}

This chapter draws on music, machine learning, and human-computer interaction to elucidate an understanding of machine learning algorithms as creative tools for music and the sonic arts. Machine learning algorithms can be applied to achieve autonomous computer generation of musical content, a goal explored from various perspectives in other chapters of this book. Our main emphasis, however, is on how machine learning algorithms support distinct human-computer interaction paradigms for many musical activities, including composition, performance, and the design of new music-making technologies. By allowing people to influence computer behaviours by providing data instead of writing program code, machine learning allows these activities to be supported and shaped by algorithmic processes. This chapter provides new ways of thinking about machine learning in creative practice for readers who are machine learning novices, experts, or somewhere in between. We begin with a brief overview of different types of machine learning algorithms, providing a friendly introduction for readers new to machine learning, and offering a complementary perspective for readers who have studied these algorithms within more conventional computer science contexts. We will then motivate a new understanding of learning algorithms as human-computer interfaces. We show that, like other interfaces, learning algorithms can be characterised by the ways their affordances intersect with goals of human users. We also argue that the nature of interaction between users and algorithms impacts the usability and usefulness of those algorithms in profound ways. This human-centred view of machine learning motivates our concluding discussion of what it means to employ machine learning as a creative tool.

\subsection{Learning More about Machine Learning and Music}
\label{ssec:learning}

A single chapter is insufficient to properly cover the use of machine learning, even within music! We will not discuss machine learning outside the context of electronic, electroacoustic, and/or experimental music creation. Readers with a more general interest in machine learning for music recommendation and analysis should investigate the literature in music information retrieval, particularly the proceedings of the International Society for Music Information Retrieval (ISMIR) conference. Those interested in algorithmic learning and re-creation of Western classical music might find David Cope's work stimulating (e.g., \cite{cope1996experiments}). Related challenges include machine learning of musical accompaniment (e.g., \cite{raphael2001synthesizing}) and expressive rendering of musical scores (e.g., the Rencon workshop, \cite{hiraga2004rencon}). Finally, readers who are new to machine learning and interested in learning more (albeit from a conventional, not arts-centric perspective) might use textbooks by \cite{witten2005data} for a practical introduction, or \cite{bishop2006pattern} for a more thorough treatment. Such resources will be helpful in addressing practical challenges, such as understanding the differences between learning algorithms, or understanding how to improve the accuracy of a given algorithm. However, any creative practitioner will also be well served by hands-on experimentation with machine learning and healthy scepticism for any official wisdom on how learning algorithms ``should" be used.

\section{Machine Learning as a Tool for Musical Interaction}

One significant advantage of machine learning is that it allows us to tackle increasingly complex musical scenarios by leveraging advances in computation and/or data resources. In this section, we begin by describing at a very high level the utility of machine learning for these types of scenarios, and by introducing the basic terms needed to describe the learning process. We then provide a practical perspective on how different families of algorithms -- each with its own approach to learning from data -- allow us to achieve common types of musical goals.

\subsection{From Executing Rules to Learning Rules} 

Creating algorithms for music making can be thought of as defining rules that will subsequently drive the behaviour of a machine. For instance, mapping rules can be defined between input data values (e.g., sounds or gestures performed by a human musician) and output values (e.g., sounds produced by the computer). Although explicitly defining these rules provides complete control over the elements in play, there are complex situations in which execution rules cannot be defined explicitly, or where defining an exhaustive set of rules would be too time consuming.

An alternative approach is to learn these rules from examples. For instance, a gesture-to-sound mapping can be defined by providing examples of input gestures, each paired with the output sound that should be produced for that gesture. Using a learning algorithm to learn these rules has several advantages. First, it can make creation feasible when the desired application is too complex to be described by analytical formulations or manual brute force design, such as when input data are high-dimensional and noisy (as is common with audio or video inputs). Second, learning algorithms are often less brittle than manually-designed rule sets; learned rules are more likely to generalise accurately to new contexts in which inputs may change (e.g., new lighting conditions for camera-based sensing, new microphone placements or acoustic environments for audio-based sensing). Third, learning rules can simply be faster than designing, writing, and debugging program code. 

\subsection{Learning from Data}

A learning algorithm builds a model from a set of training examples (the training set). This model can be used to make predictions or decisions, or to better understand the structure of the data. Its exact nature depends on the type of learning algorithm used, as we explain below. A training dataset typically consists of many example data points, each represented as a list of numerical features. A feature can be thought of as a simple, informative measurement of the raw data. For example, an audio analysis system might describe each audio example using features related to its pitch, volume, and timbre. A gesture analysis system might describe each example human pose using (x, y, z) coordinates of each hand in 3D space. Much research has considered the problem of choosing relevant features for modelling musical audio, gesture, and symbolic data (see, for example, the proceedings of the ISMIR and NIME conferences).

\subsubsection{Supervised learning}
\begin{figure}[!h]
	\centering
	\includegraphics[scale=0.4]{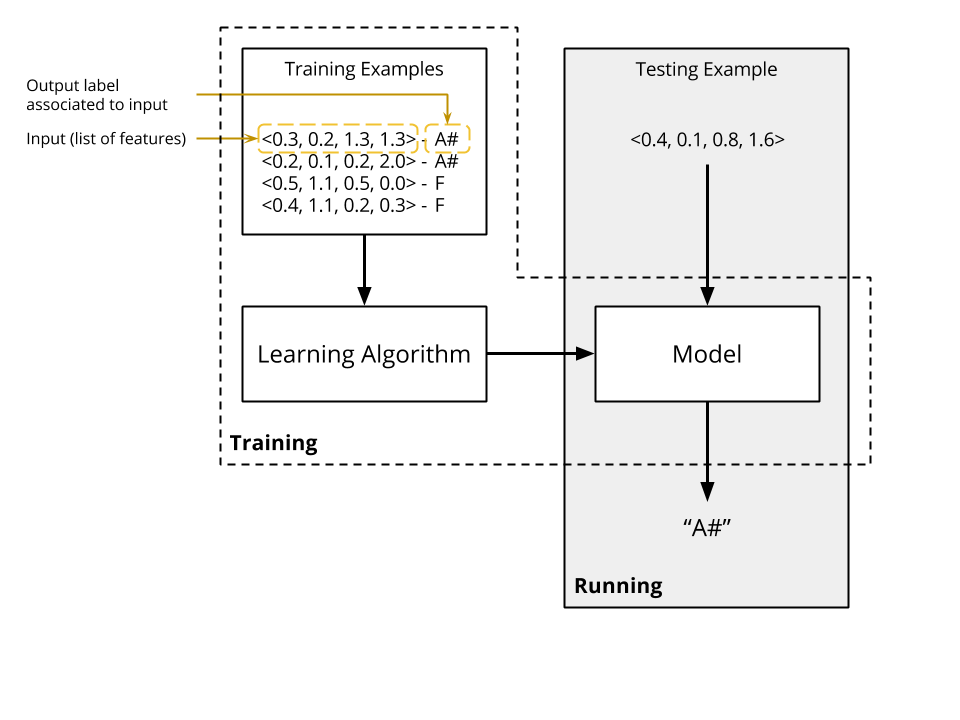}
	\caption{\footnotesize Supervised learning: The dashed line encircles the training process, in which a model is built from training examples. The shaded box denotes running of the trained model, where outputs are produced in response to new inputs.}
	\label{Fig1}
\end{figure}

\begin{figure}[!h]
	\centering
	\includegraphics[scale=0.4]{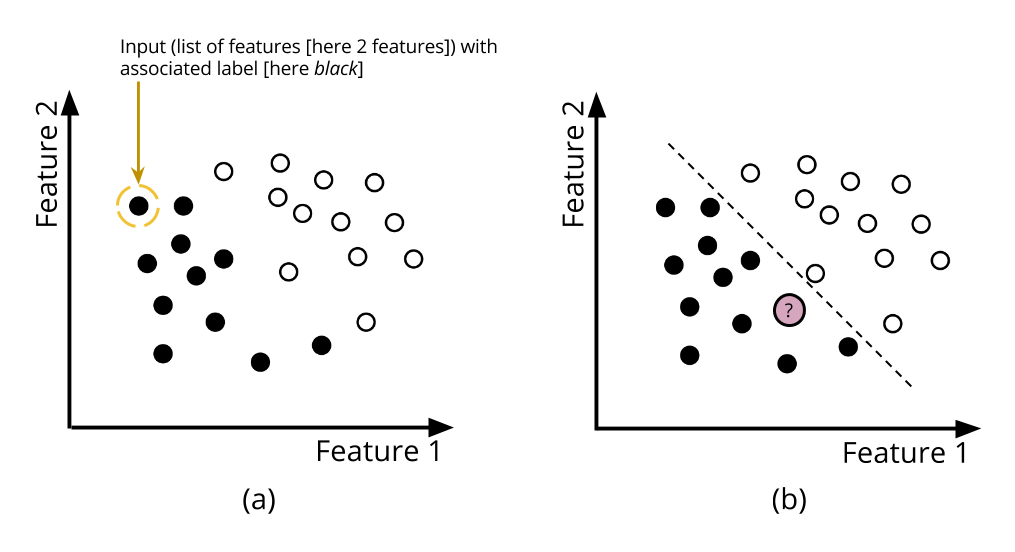}
	\caption{\footnotesize A classifier is trained on examples, which are each labelled with a class. (a) A training dataset, where each example is a point whose position is determined by the value of its two features, and whose colour is determined by its class label (black or white). (b) A classification model can be understood as a partitioning of the input space into regions corresponding to each class, separated by a decision boundary (shown here as a dashed line). When a new example is seen (corresponding to the point denoted ``?"), the classifier assigns it a label according to its position relative to the decision boundary. This new point will be labelled as black.}
	\label{Fig2}
\end{figure}

In supervised learning (Figure~\ref{Fig1}), the algorithm builds a model of the relationship between two types of data: input data (i.e., the list of features for each example) and output data (also sometimes called `labels' or `targets'). The training dataset for a supervised learning problem contains examples of input-output pairs. Once the model has been trained, it can compute new outputs in response to new inputs (Figure~\ref{Fig2}). 

For example, consider a musician who would like to associate different hand positions, captured by a video camera, to different notes played by a computer. The musician can construct a training set by recording several examples of a first hand position and labelling each one with the desired note, for instance ``A\#". She can then record examples of a second hand position, labelling each with another note, for instance ``F". The training process will learn what distinguishes an ``A\#" hand position from an ``F" hand position. After training is complete, the musician can perform these two hand positions and use the trained model to label them as ``A\#" or ``F". 

If the model outputs are categories (e.g., labels ``A\#" or ``F"), the task is typically called classification or recognition. If they are continuous values (e.g., if the model is to execute a smooth `glissando' from A\# to F as the performer's hand moves from one position to the other), the task is usually called regression.

\subsubsection{Unsupervised learning}

\begin{figure}[!h]
	\centering
	\includegraphics[scale=0.4]{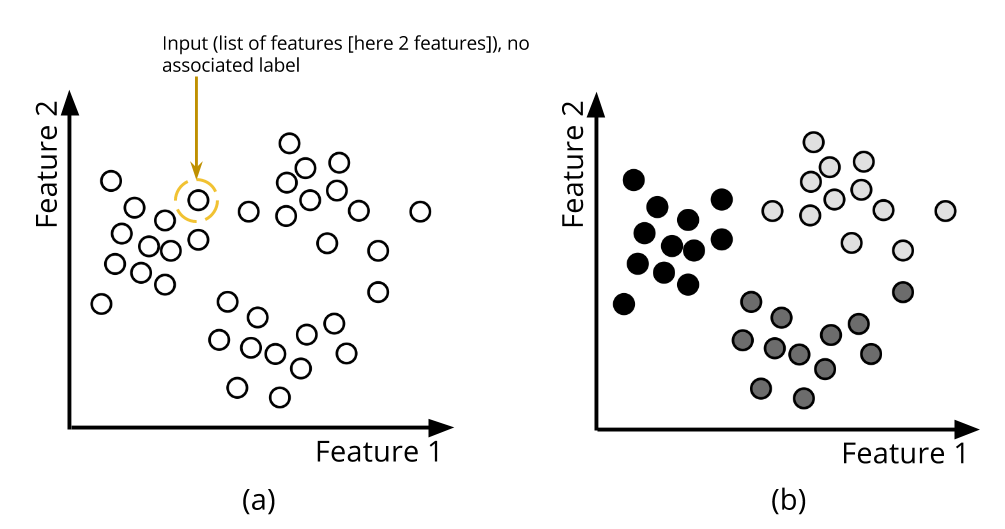}
	\caption{\footnotesize Unsupervised learning. (a) An unsupervised learning algorithm trains from a dataset where each example's feature values are known (here, the x- and y- positions of each point) but no class membership or other output information is given. (b) An unsupervised algorithm for identifying clusters of similar examples might assign these points to three clusters according to the colouring here.}
	\label{Fig3}
\end{figure}

In unsupervised learning (Figure~\ref{Fig3}), the algorithm learns the internal structure of the input data only; no corresponding output labels are provided. A musician might employ unsupervised learning simply to discover structure within the training set, for example to identify latent clusters of perceptually similar sound samples within a large sample database, or to identify common chord progressions within a database of musical scores. A musician might employ this learned structure to generate new examples similar to those in the database. Or, she might use this learned structure to provide better feature representations for further supervised learning or other processing. We return to this topic in Section~\ref{ssec:discover}. 

Consider again our musician who wants to play music by executing different hand positions in front of a video camera. However, this time she does not know beforehand how to define a suitable feature representation for hand positions. She might show the computer examples of different hand positions, without any note labels, and use an unsupervised algorithm to identify latent clusters of similar hand positions. She can then compute the most likely cluster for any new hand example and use this as an input feature in her new hand-controlled instrument.

\subsubsection{Other types of learning}

Although most uses of machine learning in music employ supervised or unsupervised learning algorithms, other algorithmic families exist. For example, in semi-supervised learning, some training examples include output labels but others do not. This approach is motivated by the fact that providing output labels for every input in the training set can be difficult and time consuming. Our hand position instrument designer might create a large unlabelled example set by moving her hand in front of the camera without providing any additional information, then select a few still images from this dataset and add labels specifying what note should be played for those hand positions. She might then apply a semi-supervised learning algorithm to build her hand position classifier, with the algorithm using the labelled examples to learn how inputs should be matched to notes, but also benefitting from the numerous unlabelled examples that provide further information about the nature of inputs it is likely to encounter. 

In reinforcement learning, an algorithm learns a strategy of action to maximise the value of some reward function. This reward could be an explicit value specified by a human user in response to each action of the algorithm. A simple example is a melody generation program that could be trained to produce ``good" melodies by a human user who presses a ``thumbs up" button (positive reward value) when he likes a melody and a ``thumbs down" (penalty or negative reward value) when he dislikes the melody. The reward could alternatively be computed, for instance using an estimate of how well the melody fits with current musical material generated by human collaborators.

\subsection{Algorithms and Musical Applications}

In the previous section, we laid out the most basic ideas behind how different learning algorithms learn from data. Numerous textbooks describe how specific algorithms actually accomplish this learning, so we refer readers interested in such details to the resources mentioned in Section~\ref{ssec:learning}. Here, though, we turn to a discussion of how these general approaches to machine learning can be matched to different types of musical goals. Specifically, we explore five goals that are relevant to many musical systems: Recognise, Map, Track, Discover New Data Representations, and Collaborate. 

\subsubsection{Recognise}

Many types of musical systems might take advantage of a computer's ability to recognise a musician's physical gestures, audio patterns, or other relevant behaviours. Musically, recognizing such behaviours enables triggering new musical events or modes.  To accomplish this, supervised learning algorithms can be used to perform classification. 

Such interaction can be used to create new gesturally controlled musical instruments. For instance, \cite{modler2000neural} describes several hand gesture controlled instruments that use neural networks to classify hand positions. Specific hand positions--measured using a sensor glove--were used to start and stop sound synthesis, excite a physical model, or select modes of control for a granular synthesis algorithm. Gesture recognition can also be used to augment performance on existing musical instruments. For instance, \cite{gillian2015kinect} use an adaptive naive Bayes classifier to recognise a set of pianist postures. Recognition of these postures during live performance allows the machine to react to gestures outside a musician's usual vocabulary. 

Other types of musical systems might take advantage of real-time machine recognition of higher-level characteristics of musical audio, such as pitch, chord, tempo, structure, or genre. Implementing such recognition systems using only computer programming and signal processing can be extremely difficult, due to the complex relationships between these semantic categories and an audio signal or its spectrum. Substantial research, including much work published at the ISMIR conference, has demonstrated the potential for classification algorithms to more accurately model these relationships. By making audio understandable to machines in human-relevant terms, such classifiers can form useful building blocks for generating musical accompaniment, live visuals, or systems that otherwise respond sensitively to human musicians.

\subsubsection{Map}

Machine learning can also be used to map input values in one domain to output values in the same or another domain. Mapping has been widely investigated in the creative domain of gestural control of sound synthesis, where properties of a musician's gesture are measured with sensors and mapped to control over sound synthesis parameters \citep{wanderley2004gestural}. Other musical applications include the generation of images from sound or vice versa \citep{fried2013cross}, and the creation of sound-to-sound mappings for audio mosaicing and timbre remapping \citep{stowell2010making}.

Designing a mapping function to generates outputs in response to inputs is a difficult task, especially for the many musical applications in which inputs and outputs are high-dimensional. The space of possible mappings is enormous, and it can be hard to know what form a mapping should take in order to satisfy the higher-level goals of the system designer, such as the creation of a musically expressive gestural controller or an aesthetically pleasing music visualisation. Supervised learning algorithms are often appropriate to this type of musical challenge, since examples of inputs and corresponding outputs can be provided together. In the common case where output values are continuous, the creation of a mapping can be achieved using regression.

Supervised learning has been used to create mappings for gestural control over sound since the early work of \cite{lee1991real} employing neural networks to gestural control of audio. By recording training examples in real-time as a person moves while listening to sound, training sets can be constructed to match a user's corporeal understanding of how gesture and sound should relate in the trained instrument \citep{fiebrink2009play,franccoise2013multimodal}.

\subsubsection{Track}

Some systems that respond to human actions do more than just recognise that an action has occurred or map from a human state onto a machine behaviour. Musical applications can benefit from the machine understanding how an action is performed, that is by tracking an action and its characteristics over time. After all, in many forms of musical activity, it is the dynamics of an action that communicate musical expression and expertise. 

Score following is one common type of tracking problem, in which the goal is to computationally align a human's musical performance--specifically, the audio signal of this performance--to a musical score. Score following allows electronic events to be synchronised to an acoustic piece whose performance is subject to expressive changes by a human performer (e.g., \cite{cont2010coupled}). Synchronising real-time sensor data to a template can also be used in the creation of new gestural controllers. For instance, work by \cite{bevilacqua2009continuous} performs real-time alignment of a gesture onto template gestures from a given vocabulary. Musically speaking, gesture alignment allows the machine to respond appropriately to the timing of a human performer, for instance playing in synchrony with the downbeat of a conductor \citep{wilson2000realtime}, or scrubbing playback position of an audio sample using the position within a gestural template \citep{bevilacqua2011online}. These types of tracking applications typically employ learning algorithms that are capable of modelling sequences of events, such as hidden Markov models or dynamic time warping. These algorithms are typically trained on user-provided examples of reference gestures or audio. 

\subsubsection{Discover New Data Representations}
\label{ssec:discover}

We can also employ learning algorithms to uncover structure within a collection of sound samples, musical scores, recordings of human motions, or other data. Unsupervised algorithms can uncover latent clusters of similar items, as well as re-map items into lower-dimensional spaces that preserve certain structural properties. These techniques are frequently used to facilitate human browsing and navigation of datasets. For example, self-organising maps have been used to create 2D interfaces for audio browsing and real-time audio playback \citep{smith2012unsupervised} in which perceptually similar sounds appear near each other in space. Supervised approaches such as metric learning can also be used to guide the learned representation to more closely match a user's perception of similarity between sounds or other data items (e.g., \citep{fried2014audioquilt}).

Discovering representations that succinctly account for the types of variation and structure present in a dataset can facilitate more accurate machine learning on subsequent tasks (e.g., recognition, mapping, following), when these representations are used as features for the data. For example, \cite{fasciani2013self} apply self-organising maps to human gesture examples in order to establish a gesture feature representation to use in mapping within a new digital musical instrument. \cite{fried2013cross} demonstrate how unsupervised deep feature learning can be applied to two domains -- e.g., gesture and audio, or music and image -- in order to subsequently build mappings between them. Such work follows a larger trend in machine learning, in which advances in machine learning of features are driving improvements in many applications involving analysis of rich media, including speech \citep{hinton2012deep}, video \citep{mobahi2009deep}, and music \cite{humphrey2012moving}.

\subsubsection{Collaborate}

Another category of musical applications involves the computer taking on a role more similar to a human musical collaborator, imbued with ``knowledge" of musical style, structure, or other properties that may be difficult to represent using explicit rules. Building an artificial musical collaborator often requires the computer to understand the sequences of human actions taken in music performance, and/or to generate appropriate sequences itself. Therefore, learning algorithms for probabilistic modelling of sequences have long been used in this context. These include Markov processes and their extensions, including hidden Markov models (HMMs) and hierarchical and variable length Markov models (e.g., \citep{ames1989markov,conklin2003music}).

Pachet's Continuator, for example, learns musical sequence patterns from a training set using variable-order Markov models \citep{pachet2003continuator}. During performance, the machine can autonomously continue sequences begun by a human musician, responding to and extending human input in a style similar to the training corpus. This type of algorithmic approach opens up several types of collaborative relationships between the algorithmic and human performers, such as mimicking the style of a famous musician or mimicking one's own style. 

Assayag's Factor Oracle \citep{assayag2004using} follows a similar approach, learning patterns from a human musician's improvisation, then improvising with the musician using the same musical material and patterns. The system is not probabilistic, but based on a syntactic analysis of the music played by the improviser. 

Machine learning can also explicitly train machine stand-ins for human performers. Derbinsky and Essl \cite{derbinsky2012exploring} use reinforcement learning to model rhythm sequences played by people in a collaborative digital ``drum circle". An intelligent agent learns the drumming style of any human performer, using its degree of match with that person's real-time performance as the reward function to guide learning. The agent can then replace that person if they leave the drum circle.  \cite{sarkar2007recognition} use dynamic Bayesian networks (a generalization of HMMs) to model tabla sequences of human drummers to facilitate collaboration between physically distant humans performing together over the Internet. Their system is trained to recognise a tabla player's current drumming pattern. The player's machine transmits only information about this pattern (which can be thought of as an extremely compressed version of the player's audio signal) to the distant collaborator, where the pattern's audio is synthesised locally.

In other work, we \citep{fiebrink2010toward} have written about the role of general-purpose supervised learning algorithms as collaborative partners for exploring new musical instrument designs. Indeed, we believe that the potential for learning algorithms to support human exploration, reflection, and discovery in the creation of new music and new technologies is an under-recognized strength, and we will detail this shift in perspective in the remainder of this chapter. 

\section{Machine Learning Algorithm as Interface}
\label{sec:interf}

Different machine learning algorithms present different assumptions about what it means to learn and how data can be used in the learning process, and it is not always clear which algorithm is best suited to a problem. Algorithms can be stymied by noisy data, by too little data, by poor feature representations. Computational perspectives on these challenges are easy to find in machine learning textbooks, as well as in the machine learning research literature. However, purely computational perspectives on what learning means and how it can be algorithmically accomplished are insufficient for understanding machine learning's potential and consequences as a tool for music creation. In this section, we describe how applied machine learning can be understood as a type of interface--not a graphical user interface, but a more fundamental relationship between humans and computers, in which a user's intentions for the computer's behaviour are mediated through a learning algorithm and through the model it produces. 

By understanding a human interacting with a machine learning algorithm as just another scenario in which a human interacts with a computer, we can bring concepts, methodologies, and value systems from human-computer interaction (HCI) to bear on applied machine learning. We begin with a consideration of the interactive affordances of learning algorithms in relationship to people creating new musical systems.

\subsection{Affordances of Learning Algorithms}

The term ``affordance" was coined by the perceptual psychologist \cite{gibsonj}, and it is now used in HCI to discuss the ways in which an object--e.g., a software program, a user interface element, a chair--can be used by a human actor.  \cite{gaver1991technology} defines affordances for an HCI readership as ``properties of the world defined with respect to people's interaction with it. Most fundamentally, affordances are properties of the world that make possible some action to an organism equipped to act." \cite{mcgrenere2000affordances}, writing about the historical use and adaptation of the concept within the HCI community, enumerate several ``fundamental properties" of an affordance. First among these is the fact that ``an affordance exists relative to the action capabilities of a particular actor." That is, an affordance is not a property of an object (or a human-computer interface) in isolation; it is a property of an object in relationship to a specific person using that object, with their specific abilities, goals, and context of use. Furthermore, ``the existence of an affordance is independent of the actor's ability to perceive it."

McGrenere and Ho show how the concept of affordances can be used to frame discussion about the usefulness and usability of human-computer interfaces. They argue that the usefulness of an interface is essentially linked to the existence of necessary affordances, whereas the usability of an interface is influenced by the ease with which a user can undertake an affordance and the ease with which they can perceive it. 
We draw on this concept to explore the usefulness and usability of machine learning in creative musical contexts. In such contexts, an affordance refers to the ways in which properties of a machine learning algorithm match the goals and abilities of a particular composer, performer, or musical instrument designer. The presence and nature of affordances thus help us to understand when and how machine learning can be useful to such users. Examining these affordances also allows us to compare alternative algorithms according to the degree to which they match particular users' goals (i.e., their usefulness), to consider the ways in which affordances are made understandable and accessible to users (i.e., their usability), and to envision new machine learning algorithms and tools that improve usefulness and usability by providing new or easier-to-access affordances. 

In this section, we examine affordances that are especially relevant to composers, performers, and instrument designers.

\subsubsection{Defining and shaping model behaviour through data}

A machine learning algorithm exposed via an appropriate software interface affords a person the ability to build a model from data, without having to describe the model explicitly in rules or code. The existence of this affordance is fundamental to the usefulness of learning algorithms for the many musical applications described above. As discussed earlier, supervised learning algorithms afford people to employ a training dataset to communicate intended relationships between different types or modalities of data, and unsupervised algorithms afford people to use data to communicate example behaviours or other properties the computer must mimic, build upon, or represent. 

Most general-purpose learning algorithms employ a few basic assumptions about the training dataset, providing opportunities for users to manipulate the nature of the trained models through changes to the data. Many algorithms assume that the training set is in some sense ``representative" of data that the model will see in the future; if there are relatively more examples of a certain ``sort" in the training set, this can be interpreted as a likelihood that the model will see relatively more examples of this sort in the future. (We are glossing over all the technical details; see \cite{bishop2006pattern} or another source for a more respectable treatment.) This property can be misused in delightful ways; for example, a composer can communicate that a model's performance on some sort of input data is more important simply by providing more examples of that sort. 

At the same time, one can imagine other types of musical goals that a person might communicate easily through example data, which are not taken advantage of by general-purpose algorithms. For instance, ``Don't generate anything that sounds like this!" or ``These are the body motions that are comfortable and visually evocative; make sure I can use these in musically interesting ways in my new instrument." The design of new learning algorithms for musical applications might thus be motivated by a desire to support new useful affordances rather than only by more conventional computational goals such as accurate and efficient modelling.

Many general-purpose machine learning algorithms perform best with a lot of training data, especially when they are applied to difficult problems, or to data with many features. However, it may be difficult for musical users to obtain or generate a large number of training examples for the problem of interest (e.g., building a recogniser for novel gestures). A variety of strategies have therefore emerged to afford the creation of models from small datasets. Properties of a model that would usually be tuned according to the data can instead be pre-defined, as is done in the Gesture Follower by \cite{bevilacqua2009continuous}. Or, models can be trained on larger, more general-purpose datasets where available and then tuned using data from an individual user via transfer learning algorithms, as demonstrated by \cite{pardo2012building} in their work building personalisable audio equalisers. Other strategies include regularisation and interactive machine learning, both discussed below. Alternatively, when a user's intention is not to build models that generalise, but rather to use models as a basis for exploring some musical or interactive space, small datasets can suffice \citep{caramiaux2015adaptive,franccoise2015motion}.

\subsubsection{Exposing relevant parameters}

Different learning algorithms expose different control parameters--configurable properties that affect the training process and the characteristics of the trained model. Many algorithm parameters are notoriously difficult to set using human intuition alone: for example, the parameters of support vector machines (SVMs) (including kernel choice, complexity parameter, others; see \cite{witten2005data}) may substantially impact the ability to accurately model a dataset.

\begin{figure}[!h]
	\centering
	\includegraphics[scale=0.4]{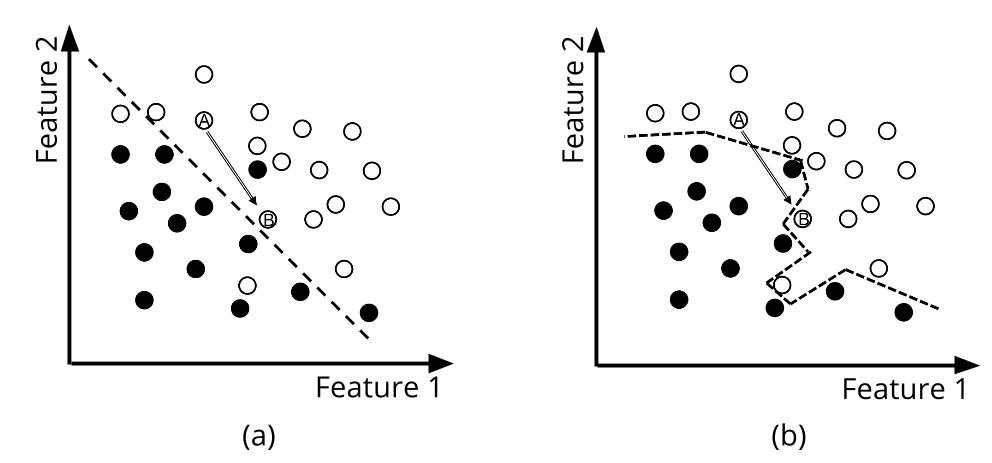}
	\caption{\footnotesize Regularisation affects boundary complexity. (a) Greater regularisation leads to smoother decision boundaries, possibly at the expense of inaccurately modelling certain training examples. (b) Less regularisation leads to more jagged decision boundaries, possibly enabling greater sensitivity to the training data. In this example, moving from A to B now crosses the decision boundary twice. }
	\label{Fig4}
\end{figure}

People using machine learning in musical contexts often care about properties of models that cannot be measured adequately with automated empirical tests, nor easily manipulated via the choice of training dataset \citep{fiebrink2011human}. Sometimes, learning algorithms can expose parameters that afford users more direct control over properties they care about. Algorithms that train using iterative optimisation often present a trade-off between training time and accuracy;  \cite{fiebrink2009play} show how a user interface can allow musicians to exercise high-level heuristic control over this trade-off when training during live performance or similarly time-sensitive contexts. As another example, some algorithms offer regularisation to control the degree to which a model fits to a training set. In classification, this can be understood as controlling the smoothness or complexity of the decision boundary (Figure~\ref{Fig4}). Regularisation can prevent a classifier's output from changing in unpredictable ways as a user smoothly changes its input (e.g., moving from A to B in Figure~\ref{Fig4}), but it can also prevent a model from accurately handling examples similar or identical to those in the training set. \cite{franccoise2015motion} describes methods for regularisation of probabilistic models of human movement that allow users to experimentally adjust models' fit, while also enabling models to be learned from small training sets. 

Some work has explored parameterising algorithms in ways that are specifically meaningful to musicians. For example, \cite{morris2008exposing} provide users with ability to manipulate a ``happy factor" and a ``jazz factor" in tuning a hidden Markov model chord generator. Pachet's Continuator \citep{pachet2003continuator} affords tuning of the extent to which a music sequence generator is governed by patterns in the training set, versus influenced by the immediate performance context (e.g., key area or volume).

\subsubsection{Modelling temporal structure}

The organisation of sound over different timescales is a defining characteristic of music. For instance, musical concepts such as phrase, form, and style can only defined with regard to patterns of sound over time. Likewise, the physical movement involved in activities such as conducting or playing an instrument entails the expressive execution of patterns of movements over time. Different learning algorithms vary in their approaches to modelling structure over time (if they model it at all); as such, models produced by different algorithms afford different types of interactions between humans, machines, and sound over time.   

When machine learning is applied in a musical context, each data point often represents a single, brief instant in time. For models produced by many general-purpose algorithms, times in the past or future are irrelevant to how the model interprets data at the current time. For example, a neural network mapping that is trained to produce a given sound in response to a performer's body pose will produce the same sound whenever that pose is encountered, regardless of the way the musician is moving through that pose. Such a model affords a tight relationship between gesture and sound over time that is similar to that which occurs when a musician performs on an acoustic instrument.

On the other hand, this type of model is incapable of instead responding to the human as if he were a conductor. A conductor moving in front of an orchestra expects very different sounds to arise from the same movements at different times in a piece, and the dynamics of her sequence of movements may be more important than her precise pose in communicating her intention. Creating this type of interaction between human and computer requires a different type of computational understanding of movement over time. For example, human motion over time can be modelled by Markov processes (e.g., \cite{bevilacqua2009continuous}). A first-order Markov model learns transition probabilities that describe the likelihood of moving to a particular next position (called a ``state"), given the previous position. Such a model affords interactions that rely on the computer remembering a sequence of past human actions and using them to anticipate the next actions. Musical sequences such as pitch and chord sequences can also be modelled by Markov processes. In such contexts, temporal modelling affords the creation of computational systems that learn notions of melodic or harmonic style from a set of examples. 

A number of strategies have been devised to simultaneously account for both low- and high-level temporal structure, as both are important to many musical phenomena. One approach is to use hierarchies of Markov models (or their ``hidden" variants), for example devoting different levels to pitch (low-level) and phrase (higher-level) of a melody \cite{weiland2005learning}. In gesture modelling, hierarchical levels can afford modelling of individual gestures as well as transitions between gestures (which also enables segmentation between gestures; see \cite{caramiaux2012segmenting}. Task-specific hierarchy definitions can also be employed; for example, \cite{franccoise2012hierarchical} decompose a physical gesture for musical control into four phases: preparation, attack, sustain and release. A Markov model is learned for each of these phases, and the system can follow a musician as he switches between these phases in order to control phase-specific sound outputs.

General-purpose algorithms that learn temporal structure in the input data usually try to become robust to variability in the data, whether temporal, spatial, or other. Variability is usually considered as noise and modelled as such. However, such variability can also be seen as a form of intentional expression by a human user, and algorithms capable of recognising and responding to variability afford the user possibility for new types of exploration and control. In speech, for instance, prosody is defined as the way a sentence is said or sung. Synthesis of expressive speech or singing voice exploits these potential variations in intonation. Similarly, musicians' timing, dynamics, and other characteristics vary across performances (and across musicians), and recent techniques allow for modelling the temporal structure of musical gesture while also identifying its expressive variations \citep{caramiaux2015adaptive}. 

\subsubsection{Running and adapting in real-time}

In many musical contexts, trained models have to respond in realtime to the actions of a performer. It is therefore often necessary to choose or customise machine learning algorithms so that the models they produce afford sufficient real-time responsiveness. 

One unavoidable challenge is that temporal models that analyse real-time sequences of inputs may have to continually respond and adapt to a sequence before the sequence has completed. A system for real-time accompaniment or control may need to generate sound while a performer plays or moves, rather than waiting for a phrase or gesture to finish. This can necessitate a change in computational approach compared to offline contexts. For example, algorithms that analyse a sequence as it unfolds in time (such as the forward inference algorithm for HMMs) may be less accurate than algorithms that have access to the full sequence (e.g., Viterbi for HMMs; see \cite{rabiner1989tutorial}). 

Training or adapting models to new data typically requires much more processing power than running a pre-trained model. However, some systems do manage to adapt in real-time, even during performance, through clever algorithm design or through strategically constructing real-time interactions to accommodate the time needed for learning. For instance, \cite{assayag2006omax} describe a machine improvisation system in which the machine learns a sequence model from human musicians' playing and uses that model to generate its own sequences. They use a factor oracle (a type of variable-order Markov chain) to model sequences, structured so that the system is capable of efficiently learning and generating sequences in real-time performance.

Even when training does not occur in a real-time performance context, the time required to train a model also impacts its interactive affordances in exciting ways, as we discuss next.

\subsection{Interactive Machine Learning}
\label{ssec:interactive}

\begin{figure}[!h]
	\centering
	\includegraphics[scale=0.6]{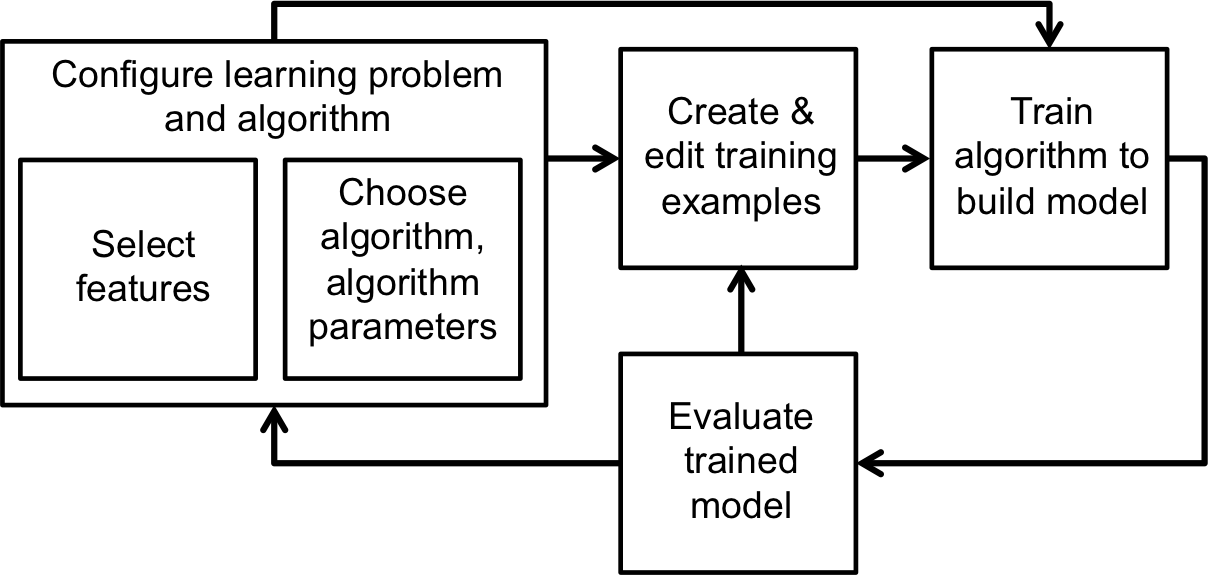}
	\caption{\footnotesize Interactive machine learning involves free-form iteration through different types of changes to the learning algorithm and data, followed by re-training and evaluating the modified model.}
	\label{Fig5}
\end{figure}

Although a user's intentions for a learning algorithm can be embedded in his or her initial choice of training data (as mentioned above), recent work shows the usefulness of enabling the user to iteratively add training examples, train the algorithm, evaluate the trained model, and edit the training examples to improve a model's performance. Such interaction is possible when training is fast enough not to disrupt a sense of interactive flow (e.g., a few seconds). Interactive machine learning is the term first used by \cite{fails2003interactive} to describe an approach in which humans can iteratively add training examples in a freeform manner until a model's quality is acceptable; it has since come to encompass a slightly broader set of techniques in which human users are engaged in a tight interaction loop of iteratively modifying data, features, or algorithm, and evaluating the resulting model (Figure~\ref{Fig5}). Fails and Olsen originally proposed this approach in the context of computational image analysis, but it has since been applied to a variety of other problems, such as webpage analysis \citep{amershi2015modeltracker}, social network group creation \citep{amershi2012regroup} and system ``debugging" \citep{groce2014you}. 

\cite{fiebrink2009meta} show how interactive machine learning can be used in music composition and instrument building. They designed a machine learning toolkit, the Wekinator, that allows people to create new digital musical instruments and other real-time systems using interactive supervised learning, with user-specified model inputs (e.g., gestural controllers, audio features) and outputs (e.g., controlling sound synthesis parameters, live visuals, etc.). Instrument builders, composers, and other people using Wekinator to create interactive music systems provide training examples using real-time demonstration (e.g., a body pose paired with a sound synthesis parameter vector to be triggered by that pose). General-purpose supervised learning algorithms for classification and regression can then learn the relationship between inputs and outputs. Users evaluate trained models by running them in real time, observing system behaviour (e.g., synthesized sounds) as they generate new inputs (e.g., body movements). Users can also iteratively modify the training examples and retrain the algorithms on the modified data. 

Interactive machine learning can allow users to easily fix many system mistakes via changes to the training data. For example, if the trained model outputs the wrong sound for a given gesture, the user can record additional examples of that gesture paired with the desired sound, then retrain. This also allows users to change the behaviour of the system over time, for example iteratively adding new classes of gestures until accuracy begins to suffer or there is no need for additional classes. Interactive machine learning can also allow people to build accurate models from very few training examples: by iteratively placing new training examples in areas of the input space that are most needed to improve model performance (e.g., near the desired decision boundaries between classes), users can allow complicated concepts to be learned more efficiently than if all training data were ``representative" of future data. We recommend readers see work by \cite{fiebrink2011human} (p. 299--303) for a more detailed discussion, and \cite{khan2011humans} for a plausible explanation combining human behavioural theory and machine learning theory.

Such interaction with algorithms can impact a musician's creative process in ways that reach beyond just producing models that are more accurate. For example, interactive machine learning using regression can become an efficient tool for exploration and for accessing unexpected relationships between human actions and machine responses. In work with composers building new gesturally controlled instruments using interactive machine learning, \cite{fiebrink2010toward} observed a useful strategy for accessing new sounds and gesture-sound relationships, while also grounding the instrument design in the composer's own ideas: composers first decided on the ``sonic and gestural boundaries of the compositional space (e.g., minimum and maximum [synthesis] parameter values and [gestural] controller positions)," then created an initial training dataset employing a few gestures and sounds at these extremes. After training neural networks on this data, the resulting continuous regression models allowed composers to move around the gesture space, discovering new sounds in between and outside the boundaries of these ``anchor" examples. 

Composers in this same study also described the value of being able to try many ideas in a short amount of time, as building instruments using machine learning was faster than writing code. Furthermore, being able to edit the training data easily meant that they could continually revise the data to reflect their own changing understanding of what the instrument should do. For example, when a composer discovered new sounds that she liked within a neural network-trained instrument, she could reinforce the presence of these sounds in the final instrument by incorporating them into new training examples. 

Allowing users to evaluate trained models using exploratory real-time experimentation also allows users to judge trained models against varied and subjective criteria, such as musicality or physical comfort, and to discover information they need to improve systems' behaviour via modifications to the training examples \citep{fiebrink2011human,zamborlin2014fluid}. Through iterations of modifying and evaluating models, users themselves learn how to effectively adjust the training data to steer the model behaviour in favourable ways. Also, iterative experimentation with models encourages users to reflect on the nature of the data they are providing. For instance, users building a gesture classification model can come to better understand the nature of a gesture and iteratively improve their skills in performing it, as \cite{caramiaux2015form} observed in workshops on embodied sonic interaction design. 

Although conceptually attractive, involving humans in tight action-feedback loops with machine learning algorithms presents some usability challenges. Learning algorithms' inevitable mistakes can be difficult for users to understand and correct. Allowing users to act on the model itself may present opportunities for them to understand how a model works, and consequently why it sometimes fails (e.g., \cite{kulesza2011oriented}). In a grey-box approach, the user has access to some parts of the internal model structure and can act on them directly. \cite{franccoise2015motion} proposed a grey-box approach for creating gesture-to-sound mappings in which users can choose between models designed for gesture recognition or gesture-to-sound mapping and between instantaneous or temporal modelling. Ultimately, making machine learning more usable by novices or experts entails helping people navigate complex relationships between data, feature representations, algorithms, and models. This is a significant challenge and a topic of on-going research across many domains (see Section~\ref{ssec:commons}).

\subsection{A Human-Centred Perspective on Machine Learning}

We have presented a human-centred view of machine learning in which learning algorithms can be understood as a particular type of interface through which people can build model functions from data. We showed that these algorithms (and the models they create) provide relevant affordances for musicians, composers, and interaction designers to achieve musical goals. Different algorithms expose particular opportunities for user control, and mechanisms for users to obtain feedback to better understand the state of a model and the effects of their own actions as they build and change a model (we also refer the reader to Bullock's chapter in this book \citep{bullock2016designing}, which provides a complementary discussion on interface design in music). 

A human-centred view demands that we consider the goals of the human(s) employing a learning algorithm to accomplish a particular task. While building an accurate model of a specific training dataset may be relevant to some people employing learning algorithms for music creation, it is likely that most people have other goals as well (or instead). These can include generating musically novel material within a rough space of styles, using learning algorithms to explore an unknown space of compositional possibilities, or building new instruments whose relationship between movement and sound ``feels" right to play, as the examples encode users' embodied practices better than systems designed by writing code \citep{fiebrink2010toward}. Using machine learning to create interactive systems can leverage cognitive properties of musical performance, such as the role of listening in planning and control of human actions \citep{caramiaux2014mapping,leman2008embodied}. 

There is ample room for future research to better understand the goals of people applying machine learning to musical tasks, and to develop new learning algorithms and software toolkits whose interactive affordances are better matched to these goals. Much more could be done to develop algorithms that are even easier to steer in useful directions using user-manipulatable parameters or training data, to develop mechanisms for people to debug machine learning systems when they do not work as intended, or to use algorithms that learn from data to scaffold human exploration of new ideas, sounds, and designs.

\section{Machine Learning as Creative Tool}

Applying a human-centred perspective to the analysis of machine learning in context, as we have presented in Section~\ref{sec:interf}, shifts the focus from technical machinery to human goals and intentions. As we argue next, this shift in perspective opens up new possibilities for understanding and better supporting creative practice.

\subsection{Roles of Machine Learning in Creative Work}

Machine learning is perhaps most obviously understood as a creative tool when a model acts as a creative agent, exhibiting human-like creative behaviours. A number of works discussed above, including those by \cite{assayag2006omax,derbinsky2012exploring,pachet2003continuator}, employ learning algorithms to generate musical material in real-time. These algorithms function as collaborators with creative agency, or even as stand-ins for other humans.

Looking more closely at such work, though, we often find that its motivations go beyond strictly replacing or augmenting human performers. \cite{assayag2006omax} write about the desire for the machine's ``stylistic reinjection" to influence human musicians: ``[A]n improvising [human] performer is informed continually by several sources, some of them involved in a complex feedback loop... The idea behind stylistic reinjection is to reify, using the computer as an external memory, [the] process of reinjecting musical figures from the past in a recombined fashion, providing an always similar but always innovative reconstruction of the past. To that extent, the virtual partner will look familiar as well as challenging."  \cite{pachet2008future} is motivated by a similar idea, that of ``reflexive interaction," in which the machine is trained to be an ``imperfect mirror" of the user. Through engaging with her reflections in this imperfect mirror, the user is helped to express ``hidden, ill-formulated ideas," to engage in creative expression without being stymied by limited expertise on an instrument, and to enter a state of Flow as described by  \cite{csikszentmihalyi1992optimal}.

Learning algorithms can also be examined with regard to their affordances as design tools, whether they are used to create new musical material or computer behaviours during performance, composition, instrument building, or other activities. Research in design and HCI suggests that practices such as sketching of incomplete ideas, exploration of the design space through rapid prototyping, and iterative refinement are essential for the creation of new technologies in any domain \citep{resnick2005design}. Furthermore, creators of new music technologies are often engaged with what design theorist Horst Rittel described as ``wicked" design problems: ill-defined problems wherein a problem ``definition" is found only by arriving at a solution. The ``specifications" for these new technologies--e.g., precisely what sort of music should they produce? how exactly should they interact with human performers?--are usually not known with certainty at the beginning of the design process, making prototyping and experimentation paramount in order for the designers to ``get the right design" as well as ``get the design right." \cite{buxton2010sketching}. 

Machine learning algorithms often naturally support these design activities \citep{fiebrink2011human}. By enabling people to instantiate new designs from data, rather than by writing code, the creation of a working prototype can be very fast. Prototypes can be iteratively refined by adjustments to the training data and algorithm parameters. By affording people the ability to communicate their goals for the system through user-supplied data, it can be more efficient to create prototypes that satisfy design criteria that are subjective, tacit, embodied, or otherwise hard to specify in code. The data can implicitly communicate the style of a machine improviser or the feel of a digital instrument. The data can alternatively act as a rough sketch of a user's ideas, allowing instantiation of a model that allows further exploration of those ideas. Thus, machine learning can allow designers to build better prototypes, to build more of them, and to use prototypes to explore a wider space of designs than building systems by programming. 

\subsection{A Comparison with Conventional Machine Learning}

We end this section by summarising the shared aims and the divergences between a ``conventional" machine learning perspective (i.e., the perspective implicit in most machine learning textbooks) and an understanding of machine learning used as a creative tool. Both perspectives are relevant to creative practitioners who want to wield learning algorithms effectively in practice, and both can inform advances in musical machine learning research.

\subsubsection{Commonalities}
\label{ssec:commons}

In both perspectives, machine learning can be seen as a powerful tool to extract information from data. Indeed, in numerous domains, machine learning algorithms are used to provide new insights into data that may otherwise be poorly understood by people. ``Big data" is driving new discoveries in scientific fields including astronomy, high-energy physics, and molecular biology \citep{jordan2011message}, and data mining influences decision-making at companies across all sectors of the economy \citep{lohr2012age}. As we have discussed in Section~\ref{ssec:discover}, discovering latent structure in musical data can support the creation of new interfaces for human exploration of that data, although the aim of these interfaces is often to scaffold new interactions rather than to simply understand the data or make decisions from it. In music, algorithms can also lend new insight into users' own data, whether by acting as an ``imperfect mirror" that invites new types of reflection on a composer's own style \citep{pachet2008future}, or alerting a cellist to the fact that her bowing articulation technique must be improved in order to allow an accurate classifier to be built \citep{fiebrink2011human}. 

Machine learning is also often used because algorithms can perform more accurately than people trying to build model functions or rule sets manually. Many learning algorithms are explicitly designed to build models that generalise well from the training data (using a rigorous definition of generalisation, and computational methods that can be demonstrated to achieve it). They easily outperform less theoretically rigorous human attempts to solve complex problems in domains such as computer vision, ontology creation, and robotics control. Music is full of similarly complex challenges, including making sense of musical audio, symbolic data (e.g., music scores), human motion or emotion, or any number of other problems involving semantic analysis of or control over high-dimensional, noisy, complex signals. As such, many challenges faced by musicians trying to build accurate models are similar to those faced by other machine learning practitioners. Machine learning practitioners may have to choose among many possible feature representations, learning algorithms, and parameters when building a model. A basic grasp of computational perspectives on machine learning is invaluable for choosing, implementing, and debugging machine learning techniques in any domain. Nevertheless, even expert intuition is often insufficient, and applied machine learning involves a great deal of experimentation with different features, algorithms, and so on. 

Needless to say, machine learning is not magic, and users in music and beyond still encounter numerous challenges for which existing learning algorithms are just inaccurate, slow, or inapplicable. Ongoing advances in fundamental machine learning research will doubtless drive advances in musical applications as well.

\subsubsection{Differences}

Unlike most conventional applications, users in musical applications often have great flexibility in their choice of training data. As discussed in Section~\ref{ssec:interactive}, users can modify the training set to steer model behaviour in useful ways: fixing mistakes, adding new classes, and so on. Iterative interaction can enable learning from smaller datasets than non-interactive learning from a fixed dataset. Musicians also often adapt their own ideas for what the computer should learn based on the outcomes of their machine learning experiments. If it turns out to be too hard to teach the computer a particular concept with a limited amount of data, for example, a musician might reasonably choose to instead model a simpler concept that is almost as useful to him in practice. (Or, if it is easier than expected to teach the machine a concept, the user may choose a more difficult concept that is even more useful to him!) Whereas most machine learning practitioners might require computational techniques and software tools to efficiently compare the accuracy of models created with different algorithms and algorithm parameterisations, creative practitioners might further benefit from tools that help them diagnose how to most effectively change the training dataset, number of classes, or other characteristics of the learning problem in order for accurate modelling to take place. 

When users are ``experts" in the problem being modelled (e.g., if they are the ones who will be performing with a new gestural controller), this also opens up new opportunities for user evaluation of models. In conventional applications, model evaluation often involves running the trained model on a test dataset (separate from the training set), or partitioning the available data many times into different versions of training and test sets (i.e., ``cross-validation"). In musical applications, though, users can often generate new data on the fly and see what the trained model does with it. This type of free-form testing affords users the ability to assess models with regard to subjective criteria: for example, ``For what types of inputs is this model's behaviour most musically interesting?" or ``What are the gestures for which this classifier is likely to fail?" \cite{fiebrink2011human} suggest that this free-form testing is invaluable to users in understanding how to improve models or deciding whether a model is ready to be used in performance, and that conventional metrics such as cross-validation may be poor indicators of a model's subjective quality.

Musical users' goals for learning algorithms sometimes differ from the conventional goal of creating a model that generalises well from the given dataset. In interactive machine learning, if the user adds new training examples to steer model behaviour in a particular direction, he may prefer models that are easily influenced by new datapoints. This can correspond to a preference for algorithms that overfit rather than those that aim to generalise from the data (e.g., models that look like Figure~\ref{Fig4}b instead of Figure~\ref{Fig4}a)--something usually viewed as undesirable in conventional machine learning \citep{fiebrink2011human}. 

In music, machine learning may be also used as a way to discover new sounds or interactive relationships, and the training data may just be a way to ground this exploration in a region of the design space a user thinks is promising. In such cases, generalisation may not be at all important, and learning fairly arbitrary models from just a few examples may be perfectly acceptable. When users employ a trained model for musical exploration, they may also seek out configurations of input data that look nothing like the data present in the training set. Conventional machine learning approaches tend not to be concerned with such ``long tail" configurations if the training data suggests they are not representative of the modelled population \citep{murphy2012machine}. However, from a creative perspective, such configurations may bring relevant new musical ideas; a model thus needs to take them into account as potential relevant inputs from the user, instead of treating their occurrence as an unlikely case that can be handled in a trivial manner. This brings important challenges in terms of model design, for example the need for fast adaptation to unexpected inputs from the user. Such challenges might also be relevant to advancements in machine learning concerned with a wider set of applications; in finance, for instance, rare and unanticipated events can have important consequences. 

\section{Discussion}

By understanding machine learning as a creative tool, used within a larger context of design practice to achieve complex and often idiosyncratic artistic goals, and in an interactive setting in which users may be able to manipulate training and testing data in dramatic ways, we can begin to imagine avenues for improving on machine learning techniques to act as better partners in creation. New algorithms and user interfaces could make it even easier to instantiate designs from data, by imposing structure on learning problems that is well-matched to the structure of particular musical tasks, as well as taking advantage of other information that users might communicate through example data. New techniques could make it even easier to explore the design space of musical technologies generated from data, to compare alternative designs, or to refine designs according to criteria that are meaningful to users.

This approach also brings scientific challenges and opportunities. A user's understanding of an algorithm's affordances can certainly be enhanced through an interactive approach to the learning phase: involving a user in teaching a model will help him to understand the model. However, the means by which a human might efficiently teach an algorithm for creative purposes remains to be explored. Moreover, the relationship between a user's perception of the quality of a model and the machine's ``perception" of its quality with regard to user-supplied inputs invites further attention. For example, \cite{akrour2014programming} show that a learning algorithm can obtain better performance by ``trusting" the competence of a user, even when that user at first makes mistakes, because the model's performance in return impacts the user's consistency. In other words, feedback between user and computer can enable both to improve their skills. 

To fully realise the potential of computers as musical tools requires taking advantage of their affordance of new interactive relationships between people and sound. Many computer music composers have written about the importance of building new human-computer relationships that transcend simple ideas of control. David Rokeby distinguishes strongly between interaction and control; his view is summarised by \cite{lippe2002real} thus: ``if performers feel they are in control of (or are capable of controlling) an environment, then they cannot be truly interacting, since control is not interaction." Robert Rowe, in his seminal book Interactive Music Systems, writes about the importance of feedback loops between human and machine in which each influences the other \citep{rowe1992interactive}. Chadabe has proposed several metaphors for musical human-machine interaction, including systems in which interaction is ``like conversing with a clever friend" (\cite{chadabe1997electric}, p.287), or ``sailing a boat on a windy day and through stormy seas" \citep{drummond2009understanding}. 

Although these composers were writing about performance-time interactions between people and machines, we argue that it is productive to characterise design-time interactions in many of the same ways. It is possible to write out detailed specifications for a new musical instrument or improvisation system, implement a system to those specifications, and be done. However, how much better to be able to discover, explore, and adapt to everything that one can learn along the way! When the computer becomes a conversation partner, or a boat rocking us in unexpected directions, we may find that the technologies we build become more useful, more musical, more interesting than our original conceptions. 

Machine learning allows us to forgo programming the machine using explicit rules, and instead makes it possible to create new technologies using more holistic strategies, using data to implicitly communicate goals and embodied practices. Machine learning allows us to create prototypes from half-baked ideas and discover behaviours we hadn't thought of, and to efficiently modify our designs in order to reflect our evolving understanding of what a system should do. In this, we can understand machine learning algorithms as more than a set of computational tools for efficiently creating accurate models from data. They can be wonderful stormy ships, conversation partners, imperfect mirrors, and co-designers, capable of influencing, surprising, and challenging us in many musical creation contexts.

\bibliographystyle{apalike}
\bibliography{biblio}

\end{document}